\documentclass[%
superscriptaddress,https://www.overleaf.com/project/6255a1fc9b0f3adbd427cc0c
preprintnumbers,
 amsmath,amssymb,
prb,
]{revtex4-2}
\usepackage{siunitx}
\usepackage{graphicx}
\usepackage{dcolumn}
\usepackage{bm}
\usepackage{hyperref}
\usepackage{gensymb}

\usepackage[
letterpaper,margin=1in]{geometry}
\usepackage{multirow}
\usepackage{xcolor}
\usepackage[section]{placeins}
\usepackage[inline]{enumitem}
\setlist[itemize]{leftmargin=*}
\setlist[enumerate]{leftmargin=*}
\setlist{noitemsep}
\usepackage{makecell}
\usepackage{lineno}

\usepackage{ragged2e}
\usepackage{array}
\newcolumntype{R}[1]{>{\raggedleft\arraybackslash}p{#1}}  
\newcolumntype{M}[1]{>{\centering\arraybackslash}m{#1}}   

\DeclareSIUnit\angstrom{\text {Å}}
\DeclareSIUnit\torr{Torr}

\begin{document}

\title{Net Magnetization and Inhomogeneous Magnetic Order in a High-T$_c$ Nickelate Superconductor}

\author{Alexander J. Grutter$^\dagger$}
\email{alexander.grutter@nist.gov}
\affiliation{ NIST Center for Neutron Research, National Institute of Standards and Technology, Gaithersburg, MD 20899, USA
}

\author{Nurul Fitriyah}
\thanks{equal contribution}
\affiliation{
Department of Physics, Faculty of Science, National University of Singapore, Singapore 117551, Singapore 
} 
\affiliation{
Department of Physics, Faculty of Science and Technology, Universitas Airlangga, Surabaya 60115, East Java, Indonesia 
}

\author{Brian B. Maranville}
\affiliation{
NIST Center for Neutron Research, National Institute of Standards and Technology, Gaithersburg, MD 20899, USA
}

\author{Saurav Prakash}
\affiliation{
Department of Physics, Faculty of Science, National University of Singapore, Singapore 117551, Singapore 
}

\author{Andreas Suter}
\affiliation{
PSI Center for Neutron and Muon Sciences CNM, 5232 Villigen PSI, Switzerland
}

\author{Jochen Stahn}
\affiliation{
PSI Center for Neutron and Muon Sciences CNM, 5232 Villigen PSI, Switzerland
}

\author{Gianluca Janka}
\affiliation{
PSI Center for Neutron and Muon Sciences CNM, 5232 Villigen PSI, Switzerland
}

\author{Xing Gao}
\affiliation{
Department of Physics, Faculty of Science, National University of Singapore, Singapore 117551, Singapore 
}

\author{King Yau Yip}
\affiliation{
Department of Physics, Faculty of Science, National University of Singapore, Singapore 117551, Singapore 
}

\author{Zaher Salman}
\affiliation{
PSI Center for Neutron and Muon Sciences CNM, 5232 Villigen PSI, Switzerland
}

\author{Thomas Prokscha}
\affiliation{
PSI Center for Neutron and Muon Sciences CNM, 5232 Villigen PSI, Switzerland
}

\author{Julie A. Borchers}
\affiliation{
NIST Center for Neutron Research, National Institute of Standards and Technology, Gaithersburg, MD 20899, USA
}

\author{A. Ariando}
\email{ariando@nus.edu.sg}
\affiliation{
Department of Physics, Faculty of Science, National University of Singapore, Singapore 117551, Singapore 
}

\date{\today}

\begin{abstract}

High-temperature and high-magnetic-field-induced re-entrant superconductivity has been discovered in the infinite-layer nickelate $\mathrm{Sm_{1-x-y} Eu_x Ca_y Ni O_2}$ (SECNO) \cite{chow2025bulk, yang2025enhanced, yang2025robust, rubi2025extreme}. Infinite-layer nickelates are the closest known analogues of high-$\mathrm{T}_c$ cuprate superconductors, yet they host distinct magnetic ground states. Using low-energy muon spin relaxation and polarized neutron reflectometry, we reveal the magnetic order in SECNO. We find that magnetic freezing occurs at a higher-temperature than in other nickelate compounds, and that a substantial net magnetization of 55 $\,\mathrm{kA}\,\mathrm{m}^{-1}$ $\pm10 \,\mathrm{kA}\,\mathrm{m}^{-1}$ emerges and remains largely unchanged across the superconducting transition. The magnetism in SECNO is disordered and nonuniform. 

\end{abstract}

\maketitle

The recently discovered superconductivity in infinite-layer (IL) and other layered square-planar nickelates opens up new avenues for investigating correlated electron superconductivity beyond the cuprates \cite{LiDiscovery2019, ZengDome2020, wang2024experimental, pan2022n5RP}. Although comparisons between nickelates and cuprates are inevitable due to their structural and electronic similarities, it remains uncertain whether the two superconducting families share a common electronic or magnetic basis mediating their pairing mechanisms \cite{kitatani2020nickelate,ferenc2023limits}. 

Similar to Cu\textsuperscript{2+} in the cuprates, Ni\textsuperscript{1+} in the IL nickelates hosts a 3d\textsuperscript{9} configuration. However, hybridization between Ni orbitals and rare-earth states complicates this picture, and a larger charge-transfer gap drives carriers into Ni 3d orbitals \cite{anisimov1999electronic, lee2004infinite, hepting2020electronic, goodge2021doping, zhang1988effective, FowlieMagnetism2022, jiang2020critical}. Bulk nickelate parent compounds exhibit no static antiferromagnetic order, in sharp contrast to the cuprates where antiferromagnetism is closely linked to superconductivity \cite{le2011intense, ortiz2022magnetic, lin2022universal, hayward2003synthesis}. Instead, bulk and thin film studies of $\mathrm{Nd_{1-x}Sr_xNiO_2}$ (NSNO), $\mathrm{Pr_{1-x}Sr_xNiO_2}$ (PSNO), $\mathrm{La_{1-x}Sr_xNiO_2}$ (LSNO), $\mathrm{NdNiO_2}$ (NNO), $\mathrm{PrNiO_2}$ (PNO), and $\mathrm{LaNiO_2}$ (LNO) support spin-glass-like states below approximately  $80\,\mathrm{K}$ \cite{FowlieMagnetism2022,saykin2025spin, lin2022universal, ortiz2022magnetic}. Magnon-like dispersive excitations have been observed in NNO thin films, but they diminish rapidly with doping \cite{lu2021magnetic}. The relationship between the magnetic ground state and superconductivity in nickelates remains unsettled, with theoretical descriptions ranging from antiferromagnetic interactions \cite{anisimov1999electronic,lee2004infinite, botana2020similarities, kapeghian2020electronic,ryee2020induced,liu2020electronic,gu2020substantial,zhang2021magnetic,zhang2020effective} to single-band Hubbard physics \cite{arovas2022hubbard,hubbard1963electron,gutzwiller1963effect}, spin-liquid behavior, spin freezing, and strongly correlated regimes with frustrated quantum criticality \cite{werner2020nickelate, leonov2020lifshitz, uematsu2018randomness, samajdar2019enhanced, choi2020fluctuation}. 


Recent developments and emerging materials highlight the importance of understanding the magnetic interactions in superconducting nickelates. $\mathrm{Sm_{1-x-y} Eu_x Ca_y Ni O_2}$ (SECNO) IL thin films show high-$\mathrm{T}_c$ superconductivity up to $40\,\mathrm{K} $\cite{chow2025bulk}. Paramagnon spectra in this system reveal a reduction in magnetic exchange coupling despite the enhanced $\mathrm{T}_c$, contrary to trends in cuprates \cite{yan2025persistent}. Intermediate Eu-doped SECNO films exhibit re-entrant superconductivity under high magnetic fields \cite{rubi2025extreme,yang2025robust}, indicating a distinctive interplay between superconductivity and magnetic order. Similar behavior in $\mathrm{Nd_{1-x}Eu_xNiO_2}$ (NENO) points to Eu dopants as a key factor \cite{vu2025unconventional}. Despite these observations, the magnetic state in SECNO has not been systematically examined. Moreover, magnetic studies of superconducting IL nickelates remain limited because superconductivity has been realized only in thin films.

Muon spin rotation/relaxation ($\mu$SR) and neutron scattering techniques are essential to closing this gap.  In the cuprates, these probes revealed correlations between magnetism and superconductivity and the evolution of spin correlations with doping \cite{anisimov1999electronic, j2006magnetic, dean2013persistence}. Using low-energy $\mu$SR (LE$\mu$SR), Fowlie {\it et al}. found that LNO films exhibit magnetic fluctuations rather than static long-range order, while Nd- and Pr-based superconducting nickelates display intrinsic magnetic states coexisting with superconductivity \cite{FowlieMagnetism2022}. 

Here, we use LE$\mu$SR and polarized neutron reflectometry (PNR) techniques to probe the magnetic ordering in superconducting $\mathrm{Sm_{0.75}Eu_{0.20}Ca_{0.05}NiO_2}$. This composition exhibits the highest reported superconducting $\mathrm{T}_c$ \cite{chow2025bulk}. We find evidence for an extended magnetic transition which begins at approximately $200\,\mathrm{K}$, significantly higher than consensus transition temperatures in other IL nickelates \cite{FowlieMagnetism2022,saykin2025spin, lin2022universal, ortiz2022magnetic}. Zero-field LE$\mu$SR indicates a highly disordered, inhomogeneous, or spin-glass-like state. PNR provides evidence of a surprisingly large net magnetization in high field, in excess of $50\,\mathrm{kA}\,\mathrm{m}^{-1}$, distributed non-uniformly within the film. This magnetization is not substantially modified across $\mathrm{T}_c$, but rather coexists with superconductivity. Thus, SECNO shows magnetic behavior similar to other IL nickelates but with higher ordering temperatures and an emergent field-induced net magnetization.


 To examine the relationship between magnetic order and high-temperature superconductivity in SECNO, we first describe the LE$\mu$SR measurements on the low-energy muon (LEM) instrument at the Swiss Muon Source. To maximize counting statistics, we measured a mosaic of four SECNO films grown on $\mathrm{NdGaO_3}$ (NGO) substrates, with a total area of 2 $\mathrm{cm^2}$, an average SECNO thickness of $7.4\,\mathrm{nm}$, and an approximately $2\,\mathrm{nm}$ thick $\mathrm{SrTiO_3}$ (STO) cap. The individual sample geometries are listed in Table \ref{tab:mosaic}. The temperature dependence of the longitudinal resistance for all samples is shown in Fig. 1(a), with a zoomed view in Fig. 1(b), demonstrating that the superconducting transition temperatures onset of all films are grouped between approximately $29\,\mathrm{K}$ and $35\,\mathrm{K}$. A representative x-ray diffraction scan in Fig.\ 1(c) shows Laue oscillations, attesting to film uniformity. For comparison, we also performed LE$\mu$SR measurements on a single STO/LCNO (6\,nm)/STO (1\,nm) film of $1\,\mathrm{cm^2}$ area with $\mathrm{T}_{c} \approx 5\,\mathrm{K}$.

\begin{figure}
    \centering
    \includegraphics[width=0.5\textwidth]{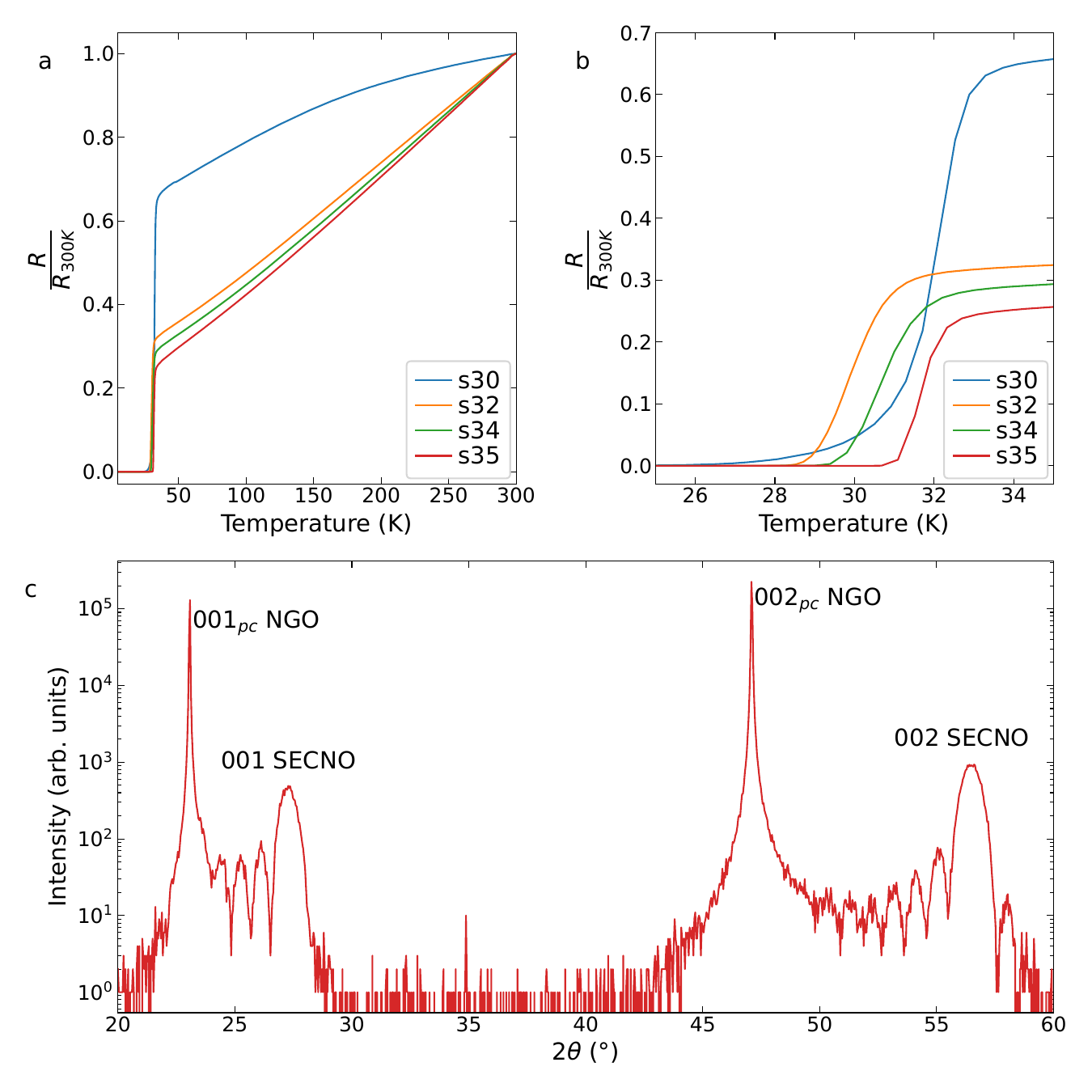}
    \caption{\justifying(a) Resistance, normalized to the value at 300 K, vs. temperature for the samples used in this study. (b) Close-up view of the temperature range encompassing the superconducting transitions of all samples, emphasizing the narrow distribution of transition temperatures. (c) x-ray diffraction measurement of s35, showing Laue fringes from the film and 00L film diffraction peaks consistent with a c-axis lattice constant of approximately 3.26\,\AA. Substrate peaks indexed using the pseudocubic (pc) unit cell.}
    \label{fig:RvTandXRD}
\end{figure}

LE$\mu$SR is a magnetic probe in which positively charged, 100\% spin-polarized muons ($\mu^+$) are implanted into a thin film. Varying the implantation energy tunes the distribution of muons within the film to provide depth-resolved information. We performed implantation simulations using the program Transport of Ions in Matter - Sputtering (TRIM.SP), Fig. 2(a), to determine the muon distribution in our SECNO mosaic \cite{biersack1984sputtering}. As in previous LE$\mu$SR studies \cite{FowlieMagnetism2022}, the IL nickelate films are relatively thin, so the fraction of muons stopping in the SECNO layer decreases above $1\,\mathrm{keV}$ as shown in Fig. 2(b). We therefore focus on measurements at $1\,\mathrm{keV}$, with additional higher-energy data provided in the supplemental information.

After implantation, the muon probes the local magnetic field environment, undergoing Larmor precession before decaying into a positron and two neutrinos. The positron is preferentially emitted along the muon spin direction, allowing the time-dependent muon spin polarization to be determined via the directional asymmetry of positron emission. We present measurements in zero applied field (ZF) and in a weak transverse magnetic field (wTF). 

In ZF measurements, the muon interacts with internal magnetic fields of the sample, with fluctuations and randomly-oriented field distributions depolarizing the muon and reducing the asymmetry over time. Following Fowlie \textit{et al. }\cite{FowlieMagnetism2022}, we describe the time-dependent ZF asymmetry data by a stretched bi-exponential decay function:
\begin{equation}
A = A_0 \left[ (1-\alpha) e^{-\lambda_F t} + \alpha e^{-(\lambda_S t)^\beta} \right] + A_{bk}
\end{equation}
where $\alpha$ is a temperature-dependent parameter characterizing the initial asymmetry of the slowly-decaying portion of the signal, and $A_{bk}$ is a temperature-independent background. The first exponential, characterized by the relaxation rate $\lambda_F$, is challenging to interpret due to interactions with muon time-of-flight effects at very short times \cite{salman2014direct}. We instead focus on the better understood slow relaxation rate, $\lambda_S$, and the stretching parameter, $\beta$. $\beta$ is sensitive to the relaxation process and magnetic field uniformity, with pure nuclear interactions in a diamagnetic system characterized by $\beta \approx 2$ (\textit{i.e.} Gaussian) while $\beta \approx 1$ for a uniform paramagnet \cite{campbell1994dynamics,hammerath2017diluted,li2016muon,KuboToyabe}. Internal field non-uniformity decreases $\beta$, with spin-glass order approaching $\beta \approx \frac{1}{3}$ \cite{biersack1984sputtering, keren1996probing}. Fowlie \textit{et al.} previously reported $\beta \approx 2$ at high temperature for NSNO, LSNO, PSNO, and LNO with $\beta \approx 0.5$ to $\beta \approx$ $1.5$ at $5\,\mathrm{K}$, indicating a range of low-temperature magnetic states \cite{FowlieMagnetism2022}.

Figures 2(c) and 2(d) show the ZF-asymmetry of SECNO and LCNO films, respectively, with no evidence of high-temperature Gaussian relaxation in either material. Rather, $\beta \approx 0.85$ for SECNO, with no statistically significant variation with temperature, as shown in Figure 2(e). A control measurement of NGO substrates shows $\beta$ to be higher across the entire temperature range ($\beta_{average} \approx 1.14$), indicating the expected uniform paramagnetism. On the other hand, $\beta$ does decrease upon cooling for LCNO films, from approximately 1.1 above $100\,\mathrm{K}$ to approximately 0.8 below $100\,\mathrm{K}$. As in Fowlie \textit{et al.}, the LCNO is grown on diamagnetic STO substrates well known for a structural phase transition at $105\,\mathrm{K}$, raising the question of extrinsic contributions to the relaxation rate.

Regardless, both SECNO and LCNO exhibit decreased initial asymmetry upon cooling. In SECNO, $\lambda_S$ increases gradually upon cooling (the peak at $100\,\mathrm{K}$ is not statistically significant, see supplemental information) while $\lambda_S$ remains flat for LCNO before increasing sharply below approximately $70\,\mathrm{K}$. Both observations are consistent with a magnetic transition, and we turn to $10\,\mathrm{mT}$ wTF measurements for a detailed examination of the temperature-dependent magnetic volume fraction. In wTF measurements, the muon spin undergoes Larmor precession, so that the asymmetry is fit to:
\begin{equation}
    A_i(t) = A_0e^{-\lambda_{\rm TF} t} \cos(B\gamma_\mu t + \phi_i)
\end{equation}
where $A_0$ is the initial asymmetry at short-time, $B\gamma_\mu$ is the Larmor precession frequency, $\phi_i$ is a detector-dependent phase factor, and $\lambda_{\rm TF}$ is the muon relaxation rate determined by the width of the magnetic field distribution and/or spin fluctuations.



Muons stopping in a magnetically ordered region of the film experience a broad local field distribution and are depolarized faster than the time resolution of the instrument, such that the muon precession effectively disappears from the measurement. Consequently, the detected muon asymmetry decreases in magnetically ordered materials, as illustrated in Fig. 2(g), which shows the wTF data for SECNO at $300\,\mathrm{K}$ and $4\,\mathrm{K}$. The magnitude of the oscillation decreases with reduced temperature, consistent with the emergence of static magnetic fields. Normalizing to a nonmagnetic state at $300\,\mathrm{K}$ and subtracting a known contribution from reflected muons \cite{suter2023low} and from muons stopping in the NGO substrate, we calculate $F_M$ within the SECNO film and STO cap:
\begin{equation}
    F_M (E,T) = 1-\frac{A_0 (E,T)}{A_0 (E,300\,\mathrm{K} )}
\end{equation}
where $F_M$ is the fraction of muons stopping in a magnetic region \textit{i.e.} near large static magnetic fields. Fig. 2(h) shows $F_M$ vs. temperature for both SECNO and LCNO. SECNO exhibits an extended magnetic transition below $200\,\mathrm{K}$. We calculate the expected $F_M$ value for the case in which every muon implanted into the SECNO layer stops in a magnetic region, based on simulations of the fraction of muons stopping in the SECNO film, NGO substrate, and STO capping layer (Fig. 2(b)) \cite{salman2014direct}. This value is plotted as a horizontal dashed line in Fig. 2(h), indicating fully magnetic SECNO below approximately $35\,\mathrm{K}$. The LCNO transition begins at lower temperature, with a gradual rise below $100\,\mathrm{K}$, followed by a sharp upturn below $50\,\mathrm{K}$. For the LCNO film, the estimated maximum F$_M$ value, consistent with 100\% magnetic LCNO and accounting for muons stopping in the STO, is plotted as a horizontal red dashed line.

\begin{figure*}
    \centering
    \includegraphics[width=1.0\textwidth]{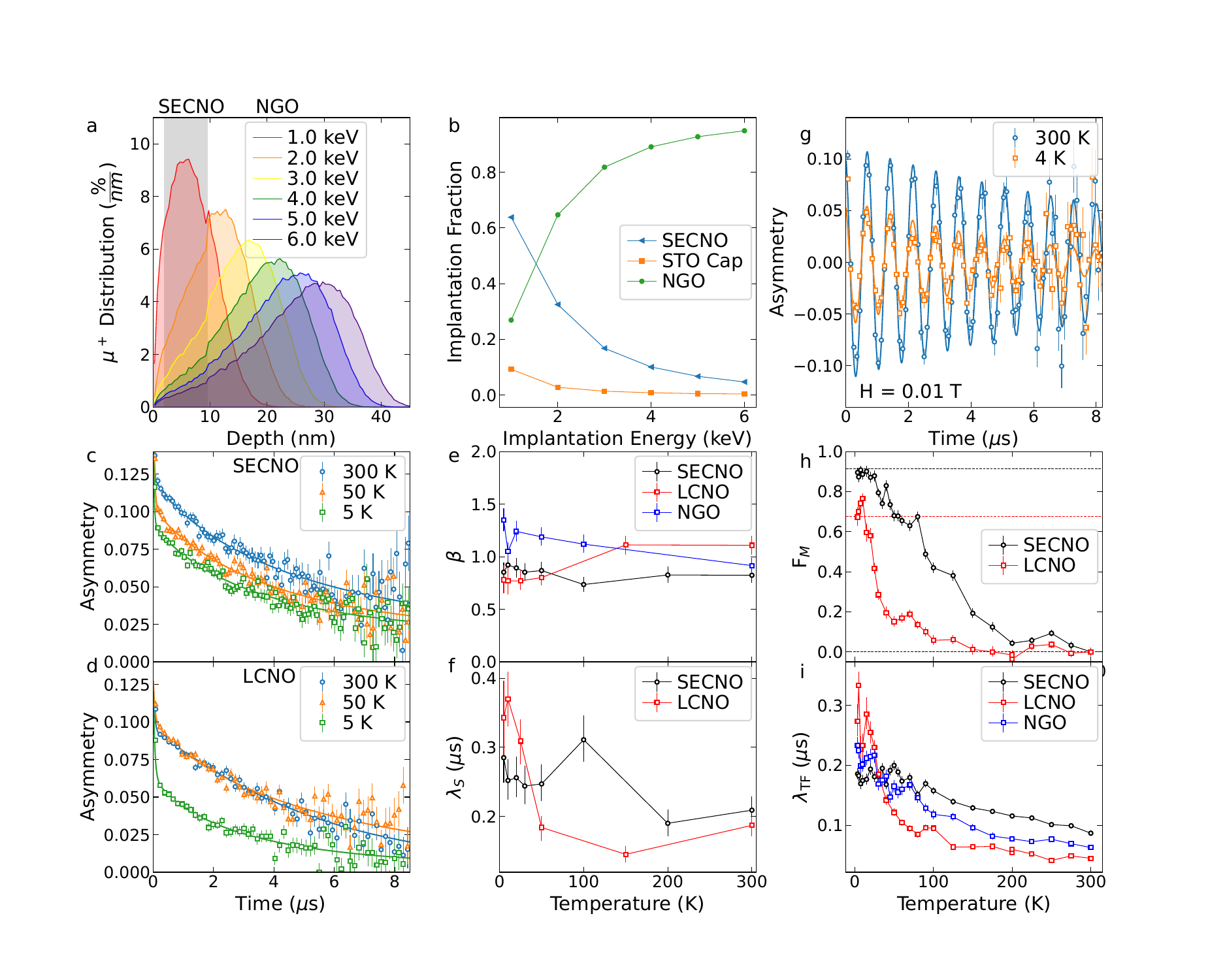}
    \caption{\justifying(a) Simulated muon implantation distribution vs. depth for the SECNO mosaic, using average thickness values. (b) Fraction of implanted muons stopping in each layer. (c) ZF LE$\mu$SR asymmetry vs. time for the SECNO mosaic at selected temperatures. (d)  ZF LE$\mu$SR asymmetry vs. time for the LCNO sample at selected temperatures. (e) Stretching parameter ($\beta$) from fits to the ZF asymmetry vs. temperature. (f) Fitted ZF muon polarization relaxation rate ($\lambda_{\rm S}$) vs. temperature. (g) Representative wTF LE$\mu$SR asymmetry vs. time for the SECNO mosaic in 10 mT applied field at $300\,\mathrm{K}$ and $4\,\mathrm{K}$, alongside theoretical fits. (h) F$_M$ vs. temperature for SECNO and LCNO. (i) wTF muon polarization relaxation rate for SECNO, LCNO, and NGO substrate vs. temperature. Error bars represent $\pm 1$ standard deviation.} 
    \label{fig:ZFmuSR}
\end{figure*}

Above $50\,\mathrm{K}$, where a significant fraction of SECNO is non-magnetic, the wTF relaxation rate (Fig. 2(i)) of SECNO exceeds that of bare NGO substrates, indicating either increased magnetic fluctuations or a broader local field distribution in agreement with the ZF data. Below approximately $50\,\mathrm{K}$, the SECNO and NGO relaxation rates converge, likely because any muons implanted into the SECNO layer immediately depolarize and do not contribute to the signal. Much like $F_M$, $\lambda_{\rm TF}$ of SECNO varies continuously with temperature rather than showing a sharp transition. The LCNO film, on the other hand, exhibits a lower depolarization rate above $70\,\mathrm{K}$, with a sharp increase from $70\,\mathrm{K}$ to $5\,\mathrm{K}$.


Thus ZF and wTF LE$\mu$SR measurements indicate inhomogeneous magnetic order in SECNO, with disordered magnetic freezing beginning near $200\,\mathrm{K}$ and plateauing by approximately $35\,\mathrm{K}$. Recent reports of re-entrant high-field superconductivity in SECNO films suggest that Eu-based ferromagnetism may play an important role \cite{yang2025robust}. Thus, understanding the physical distribution of the magnetic order is critical. We therefore obtained the depth-resolved net magnetization distribution using PNR on SECNO sample s34 using the AMOR instrument at the Swiss Spallation Neutron Source \cite{stahn2016focusing,klauser2018selene}.

\begin{figure}
    \centering
    \includegraphics[width=0.5\textwidth]{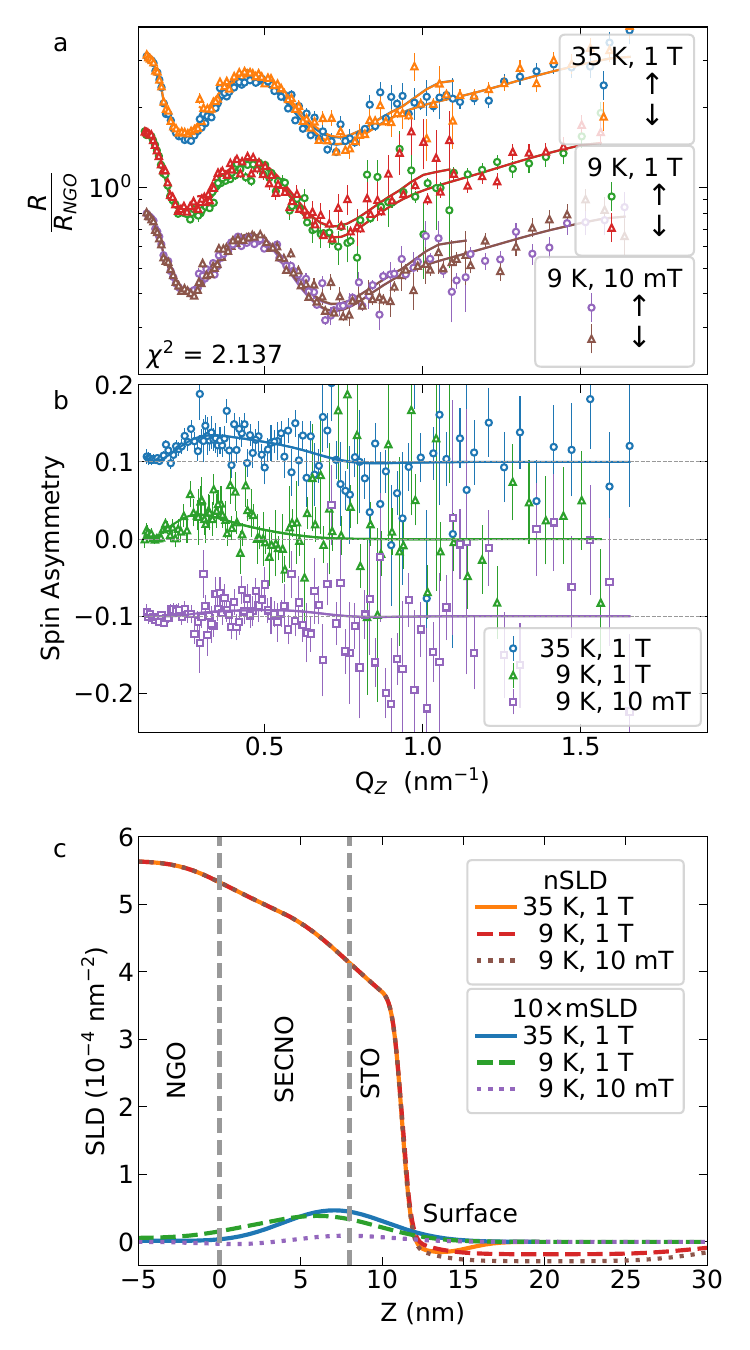}
    \caption{\justifying(a) Spin-dependent neutron reflectivity, normalized by theoretical substrate reflectivity, vs. $Q_Z$, alongside theoretical fits. Curves offset for visual clarity. (b) Spin asymmetry vs. $Q_Z$ calculated from the data in (a), alongside theoretical curves. Curves offset for visual clarity. (c) Best-fit nuclear and magnetic neutron scattering length densities vs. distance from the NGO/SECNO interface ($Z$), based on model with two distinct magnetic regions in the SECNO. Error bars represent $\pm1$ standard deviation.}
    \label{fig:PNR}
\end{figure}

Incident neutrons were spin-polarized parallel or antiparallel to an applied in-plane magnetic field, and the specular reflectivity was measured as a function of the scattering vector, $Q_Z$, along the film normal direction. The spin-dependent reflectivities ($R^\uparrow$ and $R^\downarrow$) are determined by the depth profile of the nuclear scattering length density (nSLD), dependent on the composition and density, and magnetic scattering length density (mSLD), dependent on the net in-plane magnetization. PNR measurements were taken at $35\,\mathrm{K}$ in a $1\,\mathrm{T}$ applied magnetic field, and at $9\,\mathrm{K}$ in $1\,\mathrm{T}$ and $10\,\mathrm{mT}$ applied fields. Superconductivity in $\mathrm{Sm_{0.75}Eu_{0.20}Ca_{0.05}NiO_2}$ persists well above $1\,\mathrm{T}$, such that both $9\,\mathrm{K}$ measurements are in the superconducting state \cite{yang2025enhanced}.

Figure 3(a) shows the spin-dependent neutron reflectivities, normalized by the theoretical reflectivity of a bare substrate, alongside theoretical fits. The magnetic splitting of the reflectivities is better visualized in Fig. 3(b), which plots the spin asymmetry (SA):
\begin{equation}
    \mathrm{SA} = \frac{R^\uparrow - R^\downarrow}{R^\uparrow + R^\downarrow}
\end{equation}
We observe nonzero SA, indicating a net magnetization, in both $1\,\mathrm{T}$ measurements. The best-fit nSLD and mSLD models, shown in Fig. 3(c), were obtained by fitting all three datasets with an identical nSLD structure except for a variable surface adsorbate layer. This adsorbate is similar across the two $9\,\mathrm{K}$ measurements and largely disappears at $35\,\mathrm{K}$, consistent with a wide range of PNR measurements in the literature \cite{aboljadayel2023determining}. The nSLD of SECNO, $5.03\cdot 10^{-4}\,\mathrm{nm^{-2}}$, is very near the expected value of $5.02\cdot10^{-4}\,\mathrm{nm^{-2}}$, indicating that a high-density film with a bulk-like structure has been achieved. Nevertheless, properly fitting the reflectivity from both spin states requires that the SECNO be treated as two magnetically distinct layers. 

Both $1\,\mathrm{T}$ measurements suggest that the net magnetization is concentrated towards the top of the film, being suppressed near the film/substrate interface. We find net magnetizations in the upper region of the film of $55\, \pm 10\,\mathrm{kA}\,\mathrm{m}^{-1}$ and $40\, \pm 10\,\mathrm{kA}\,\mathrm{m}^{-1}$ at $35\,$ and $9\,\mathrm{K}$, respectively. These values agree within uncertainty, and suggest that the net magnetization of the sample is not significantly modified by entry into the superconducting state. At $10\,\mathrm{mT}$, the net magnetizations in the top and bottom portions of the film are $-5 \, \pm 4 \,\mathrm{kA}\,\mathrm{m}^{-1}$ and $14.7\, \pm 7.7\,\mathrm{kA}\,\mathrm{m}^{-1}$, both statistically indistinguishable from zero. Models which describe the scattering using either a single magnetic layer in SECNO, or by confining all net magnetization to paramagnetic NGO, increased the reduced $\chi^2$ from 2.137 to 2.28 and 2.55, respectively (see supplemental information).

The nonuniform net magnetization in our SECNO films is consistent with previous reports of superconducting nickelate films which note reduced crystal quality with distance from the film/substrate interface \cite{ferenc2023limits,osada2021LSNO,Lee2023,ikeda2013improved,goodge2023resolving, he2020polarity,lee2020aspects,krieger2022synthesis,parzyck2024synthesis}. Despite the high-density indicated by PNR, the roughness of the $\mathrm{SECNO\,|\,STO \,}$ cap interface  is high, potentially indicating a gradient in oxygen content or defect density. Such a gradient has significant implications for the spin-state of Eu, which varies between nonmagnetic and high-spin states for $\mathrm{Eu}^{3+}$ and $\mathrm{Eu}^{2+}$, respectively \cite{mairoser2015high, PhysRevB.94.195101}. To fully understand this phenomenon, further studies of the magnetization depth profile at higher Eu concentrations supporting the re-entrant state are required.


LE$\mu$SR and PNR measurements reveal coexistence between disordered, inhomogeneous magnetism and a net magnetization with superconductivity in SECNO. LE$\mu$SR is consistent with the disordered state observed in other superconducting IL nickelates, with a gradual transition to higher $F_M$ as the spins freeze at lower temperatures and become ``static'' relative to the muon timescale. The onset temperature of magnetic freezing in SECNO is higher than in previous\ compounds, exceeding LCNO by $100\,\mathrm{K}$. The LE$\mu$SR depolarization rate and stretching parameters of SECNO indicate a broad distribution of local fields at high temperature, relative to LCNO, possibly related to  the Eu dopants seen as key to much of the unusual physics. We observe a substantial net magnetization in SECNO even at comparatively low Eu-doping, and the stable magnetization profile across the superconducting $\mathrm{T}_c$ suggests coexistence between ferromagnetic order and superconductivity. Our observations have important implications for the underlying pairing mechanism and recently observed re-entrant superconductivity \cite{rubi2025extreme,yang2025robust}.

\section{Experimental Methods}

\subsection{Sample Synthesis}
A typical solid-state reaction was employed to fabricate ceramic targets with nominal composition $\mathrm{Sm_{1-x-y} Eu_x Ca_y Ni O_3}$ using high-purity powders of $\mathrm{Sm_2 O_3}$  (99.999\,\%, Sigma-Aldrich), $\mathrm{Eu_2 O_3}$ (99.999\,\%, Sigma-Aldrich), $\mathrm{Ca C O_3}$ (99.995\,\%, Sigma-Aldrich), and NiO (99.99\,\%, Sigma-Aldrich). The combined powders were completely mixed and sintered in air at $1200\,\mathrm{\degree C}$, $1300\,\mathrm{\degree C}$, and $1400 \,\mathrm{\degree C}$ for 12\,h, 15\,h, and 18\,h, respectively, with regrinding in between each stage. After the fine powders were fully sintered, they were compacted into disk-shaped pellets. Thin films were grown on NGO (110) substrates by pulsed laser deposition (PLD) using a 248\,nm KrF excimer laser. All depositions were carried out at $700 \,\mathrm{\degree C}$ under an oxygen partial pressure of 26.7\,kPa, with a laser energy density of $2.4\,\mathrm{J\,cm^{-2}}$ at the target surface. To obtain the infinite-layer phase, the as-grown films underwent a process called a topotactic chemical reduction in the presence of $\mathrm{CaH_2}$  powder by heating at $\mathrm{280 \,\degree C \mbox{ to } 310 \,\degree C}$ for 4\,h. 

\subsection{Electrical Transport}
Quantum Design's Physical Property Measurement System was used for resistivity measurements down to 2 K. The electrodes for electrical transport measurements were prepared by ultrasonic wire bonding using Al wires.

\subsection{X-ray Diffraction}
The X-ray Diffraction and Development (XDD) beamline at Singapore Synchrotron Light Source (SSLS) was used for the measurement, with an X-ray wavelength equal to 1.5404\,Å.

\subsection{Low-Energy Muon Spin Relaxation}

Low-energy muon spin relaxation measurements were performed using the low-energy muon (LEM) instrument at the Swiss Muon Source of the Paul Scherrer Institute \cite{morenzoni2002implantation, prokscha2008new}. Experiments on $\mathrm{La_{1-x}Ca_xNiO_2}$ (LCNO) were performed on a single $10\,\mathrm{mm}\times10 \,\mathrm{mm}$ sample with the geometry STO/LCNO (6 nm)/STO (1 nm). The NGO and SECNO measurements were performed on sample mosaics with matching geometric configuration and a combined cross sectional area of $2 \,\mathrm{cm^2}$. The samples were mounted with silver paint on a nickel-coated aluminum plate to minimize background effects. The SECNO mosaic had an average geometry of NGO/SECNO (7.4 nm)/STO (2 nm) with the individual sample geometries listed in Table \ref{tab:mosaic}. Samples were mounted in a helium flow cryostat (CryoVac, Konti), which is capable of maintaining the sample stage temperature within $\pm 0.1\,\mathrm{K}$ of the target temperature. Fully polarized muons ($\mu^+$) are accelerated to variable energies to implant them at different depths. Implantation profiles modeled using TRIM.SP are used to select energies probing different depths \cite{morenzoni2002implantation}. While this work focuses primarily on 1\,keV implantation energies, we used energies spanning a range from 1\,keV to 18\,keV. Beam transport settings were set to 12.0\,kV. Weak transverse field measurements were performed initially at 200\,K and then cooling to the minimum temperature before returning to 200\,K and increasing the temperature to 300\,K. In general, zero-field measurements were performed by degaussing the magnet at high temperature (300\,K) and then cooling to 5\,K.

\begin{table*}[ht!]
    \centering
    \begin{tabular}
{|M{0.14\columnwidth}|M{0.14\columnwidth}|M{0.25\columnwidth}|M{0.30\columnwidth}|M{0.10\columnwidth}|}    
\hline
\textbf{Sample ID} & \textbf{Size ($\mathrm{mm^2})$} & \textbf{SECNO Thickness (nm)} & \textbf{STO Cap Thickness (nm)} & \textbf{$\mathrm{T_c}$ (K)}\\
\hline
s30 & 5$\times$5 & 8.0 & 1.0 & 26.0\\
\hline
s32 & 5$\times$5 & 8.0 & 1.0 & 28.7\\
\hline
s34 & 10$\times$10 & 7.8 & 3.0 & 29.3\\
\hline
s35 & 10$\times$5 & 6.0 & 1.0 & 31.0\\
\hline
    \end{tabular}
    \caption{Table of properties for each heterostructure in the SECNO mosaic.}
    \label{tab:mosaic}
\end{table*}

\subsection{Neutron Reflectometry}
All PNR measurements were performed using the AMOR instrument at the Swiss Spallation Neutron Source of the Paul Scherrer Institute \cite{stahn2016focusing,klauser2018selene}.

Sample s34 of our SECNO mosaic was mounted with silver paint to a standard AMOR 10\,mm by 10\,mm aluminum sample holder in an argon-filled glove box to minimize oxidation.  The sample was transported to the instrument in an argon-filled box and exposed to air for less than 1 minute before being mounted in an evacuated chamber for the measurements.  Neutron-absorbing masks surrounding the sample mount minimized background.

The AMOR reflectometer is multiplexed in both angle and wavelength, using an area detector to measure a range of scattered angles and time-stamping to measure the wavelength range.  We measured with the instrument in two different configurations: one lower-angle for low-$Q$ covering the critical angle, and another at a higher average incident angle extending out to approximately $Q = 1.5\,\textrm{nm}^{-1}$ to cover the next thickness oscillation.  The accepted angular divergence at each setting is approximately $1.2\degree$ (in the scattering plane), with wavelengths from 0.3\,nm to 1.05\,nm.

Data were reduced using the EOS python package \cite{stahn_amor_eos}.

Neutron reflectometry data were fit to a single coupled model for all three datasets in which the nuclear scattering potential (nSLD) and roughness values of all layers were constrained to be identical except for a surface adsorbate layer, which was allowed to vary. The magnetic potential (mSLD) was allowed to vary within the SECNO layer, as two sub-layers with identical nSLD but different mSLD. The NGO magnetization was constrained to follow expected paramagnetic behavior, being linear in field and proportional to 1/T. Uncertainty estimations were obtained using a Markov-chain Monte Carlo (MCMC) simulation with $10^6$ samples after convergence, as implemented in the DREAM algorithm in the bumps python package \cite{Refl1D, kirby2012phase}. Data were fit using Refl1D \cite{Refl1D}.

\begin{acknowledgments}

Certain commercial equipment, instruments, software, or materials are identified in this paper in order to specify the experimental procedure adequately. Such identifications are not intended to imply recommendation or endorsement by NIST, nor it is intended to imply that the materials or equipment identified are necessarily the best available for the purpose. Unless otherwise noted, NIST work was funded solely by the United States Government. We also acknowledge the support from the Ministry of Education (MOE), Singapore, through the Tier-2 Academic Research Fund (AcRF, Grant No. MOE-T2EP50123-0013), the SUSTech-NUS Joint Research Program, and the MOE Tier-3 Grant (MOE-MOET32023-0003) titled “Quantum Geometric Advantage”.
This work is based upon experiments performed on Amor at the
Swiss spallation neutron source SINQ, and on LEM at the Swiss muon source, both at the Paul Scherrer Institute, Villigen, Switzerland. Partial support of AJG and BBM was provided by  the Center for High Resolution Neutron Scattering, a partnership between the National Institute of Standards and Technology and the National Science Foundation under Agreement No. DMR-2010792.

\end{acknowledgments}


\section{Supplemental Information}

\renewcommand{\thetable}{S\arabic{table}}
\setcounter{table}{0}
\renewcommand{\thefigure}{S\arabic{figure}}
\setcounter{figure}{0}

\subsection{Additional Low-energy Muon Spin Relaxation (LE$\mu$SR) Analysis and Information}

\subsubsection{$\mathrm{LCNO}$ implantation profiles and measurement}

\begin{figure*}
    \centering
    \includegraphics[width=0.9\textwidth]{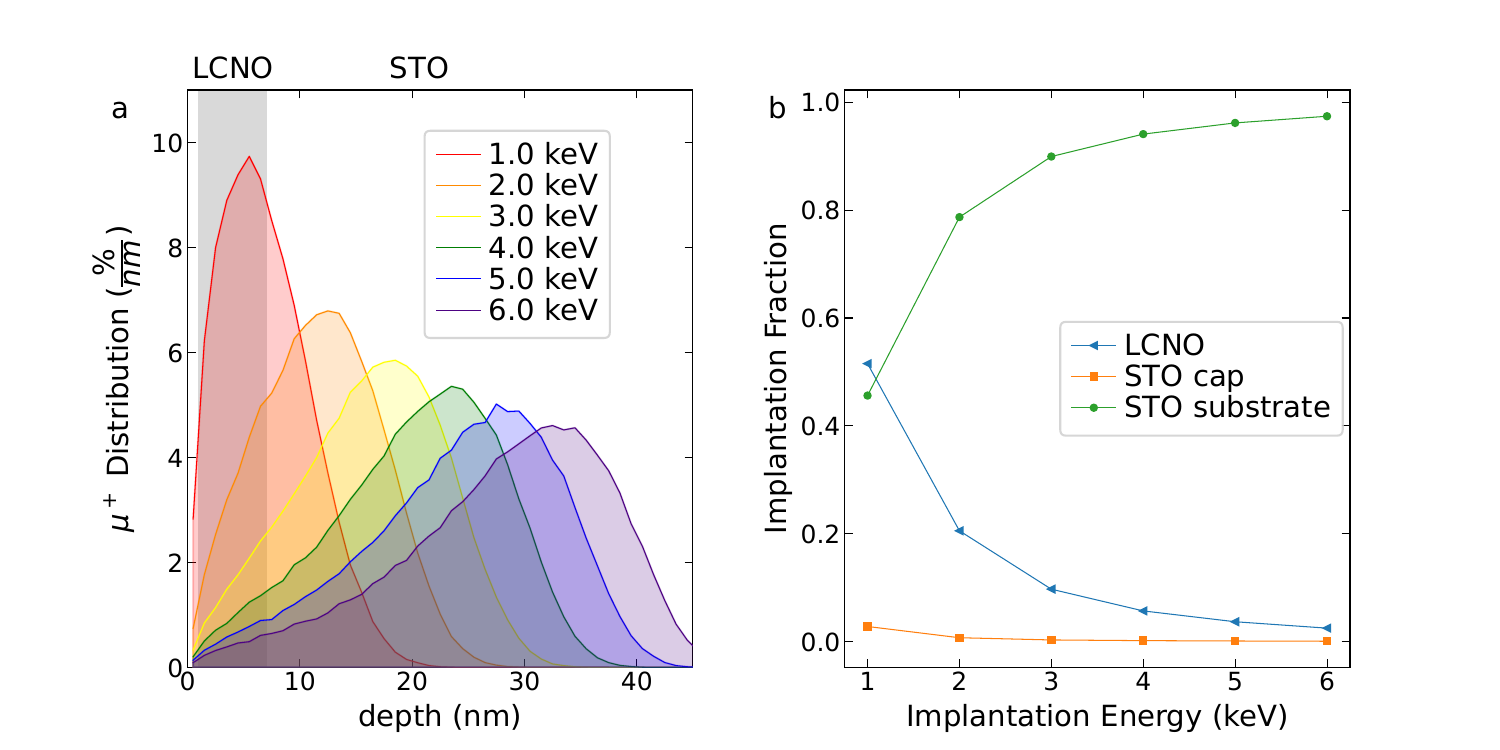}
    \caption{\justifying(a) Simulated muon implantation distribution vs. depth for the LCNO sample. (b) Fraction of implanted muons stopping in each layer.}
    \label{fig:LCNO_TRIM}
\end{figure*}

Measurements on the LCNO were performed on a single sample with cross sectional area 10 mm $\times$ 10 mm. The LCNO was grown on (001)-oriented STO and capped with approximately 1 nm of STO. Based on this geometry, we used TRIM.SP to generate the expected distribution of muon depths shown in Figure \ref{fig:LCNO_TRIM} \cite{biersack1984sputtering}. Based on these simulations, an implantation energy of 1 keV was selected. At this energy, we expect 54\% of implanted muons to stop in the LCNO, while 9\% and 37\% stop in the STO cap and substrate, respectively.

Note that at 1 keV, a significant fraction of muons are reflected from the sample surface, contributing a background muon asymmetry to the measurement. Figure \ref{fig:LCNO_TRIM}(b) shows the fraction of successfully implanted muons stopping in each layer as a function of implantation energy, but does not take into consideration the reflected muons. Instead, the background asymmetry associated with reflected muons, which has been characterized in detail on the LEM instrument, has been subtracted out in plots of F$_M$\cite{suter2023low}.

\subsubsection{Characterization of $\mathrm{NdGaO_3}$ substrates}

To our knowledge, there are no examples in the literature of LE$\mu$SR measurements being performed on films grown on NdGaO$_3$ (NGO) substrates. As the example of muonium formation in STO highlights, it is critical to understand the behavior of muons stopping in the NGO, so that their behavior can be accounted for in the analysis of the SECNO film. We therefore performed both weak transverse field (wTF) and zero-field (ZF) measurements on a mosaic of bare NGO substrates. The substrates were arranged with a cross-sectional area equal to SECNO mosaic, with a matching geometry and position on the sample holder.

\begin{figure*}
    \centering
    \includegraphics[width=0.9\textwidth]{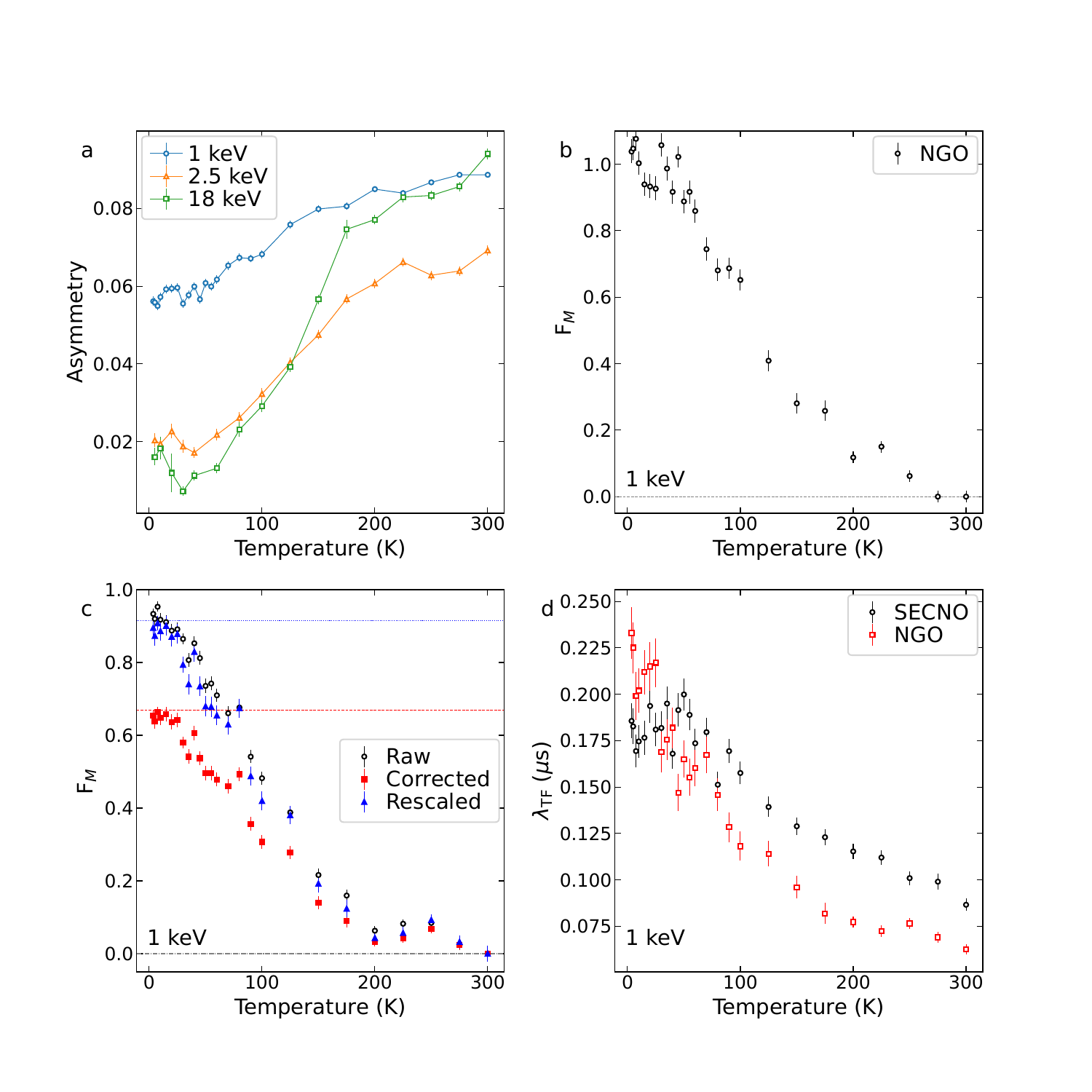}
    \caption{\justifying(a) Temperature-dependent LE$\mu$SR asymmetry from NGO substrates at implantation energies of 1 keV, 2.5 keV, and 18 keV. (b) Calculated F$_M$ for NGO for 1 keV muons. (c) Raw F$_M$ from the SECNO mosaic and the corrected F$_M$ after subtracting the contribution from muons stopping in the NGO substrate. The rescaled curve shown in the main text is obtained by dividing the corrected data by the fraction of muons stopping in the STO and SECNO. (d) Muon polarization relaxation rate vs. temperature for SECNO and NGO mosaics at 1 keV implantation energy. Error bars represent $\pm$1 standard deviation.}
    \label{fig:NGO_MuSR}
\end{figure*}

Temperature-dependent weak transverse-field measurements were performed at 1 keV in an applied field of 10 mT, as shown in Fig. \ref{fig:NGO_MuSR}. Additional measurements were performed at higher implantation energies (2.5 keV and 18 keV) on a single 10 mm $\times$ 10 mm substrate with a collimator reducing the beam area. The asymmetry values extracted from the higher-energy, lower-area measurements do not yield directly comparable asymmetry values but are useful in examining temperature trends at different implantation depths. We present these measurements in Fig. \ref{fig:NGO_MuSR}(a). Figure \ref{fig:NGO_MuSR}(b) shows trends in F$_M$ vs. temperature, with a continuous increase upon cooling likely related to the slow fluctuation rate of the large paramagnetic Nd moment. To calculate F$_M$ for SECNO, we scale the curve in \ref{fig:NGO_MuSR}(b) by the fraction of muons stopping in the NGO at 1 keV (27\%) and subtract the result from the raw F$_M$ vs. temperature curve. The raw and corrected F$_M$ curves for SECNO are shown in \ref{fig:NGO_MuSR}(c) alongside a red dashed line representing the expected (corrected) F$_M$ value associated with full magnetically ordered SECNO. This value accounts for the fraction (64\%) of implanted muons stopping in the SECNO as well as a small contribution from muonium formation in STO - 9\% of muons with effective F$_M$ of approximately 0.3 at 1 keV \cite{salman2014direct}. To obtain the dataset shown in Fig. \ref{fig:ZFmuSR}(h), we rescale the curve by the calculated fraction of muons stopping in SECNO and STO, yielding the blue ``Rescaled'' curve and the revised blue dotted line corresponding to fully magnetic SECNO in \ref{fig:NGO_MuSR}(c). Figure \ref{fig:NGO_MuSR}(d) compares the muon polarization relaxation rates for SECNO and NGO at a 1 keV implantation energy. 

\subsubsection{$\mathrm{SECNO}$ and $\mathrm{LCNO}$ wTF LE$\mu$SR energy dependence}

\begin{figure*}
    \centering
    \includegraphics[width=0.9\textwidth]{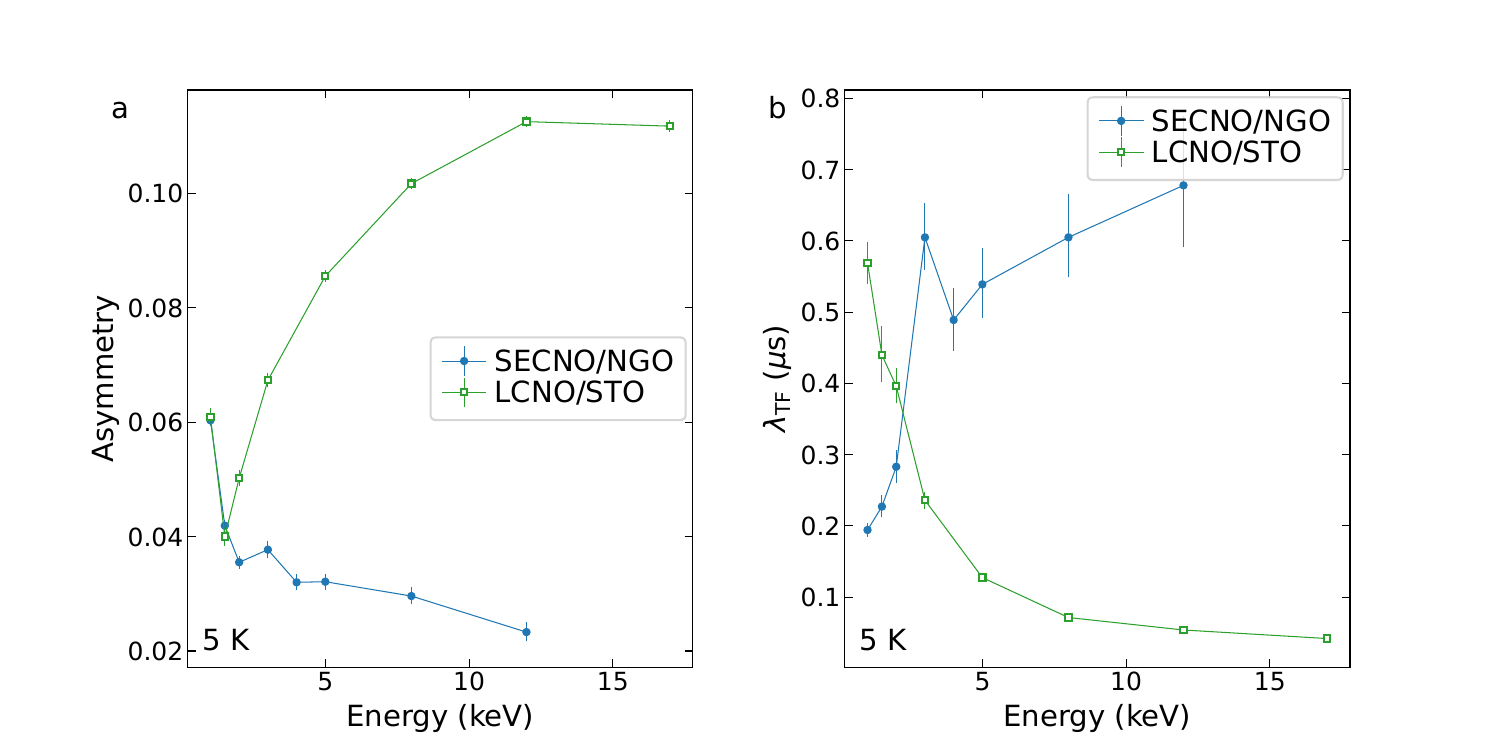}
    \caption{\justifying(a) Initial muon polarization asymmetry vs. implantation energy at 5 K in a 10 mT weak transverse field, for SECNO mosaic and LCNO sample. (b) Muon polarization relaxation rate vs. implantation energy at 5 K in a 10 mT weak transverse field for SECNO mosaic and LCNO sample.}
    \label{fig:EnergyDependence}
\end{figure*}

Figure \ref{fig:EnergyDependence} shows the wTF 5 K asymmetry vs. implantation energy at an applied field of 10 mT. As expected from the asymmetry vs. temperature measurements shown in Fig. \ref{fig:NGO_MuSR}(a), the low-temperature asymmetry rapidly decreases at higher energies in which most muons stop in the NGO. At the same time, $\lambda_{\rm TF}$ increases (Fig. \ref{fig:EnergyDependence}(b)). On the other hand, the asymmetry rises and $\lambda_{\rm TF}$ decreases at higher-energies for LCNO/STO where most muons stop in the STO.

\subsubsection{Additional zero-field (ZF) LE$\mu$SR data and analysis}

\begin{figure*}
    \centering
    \includegraphics[width=0.9\textwidth]{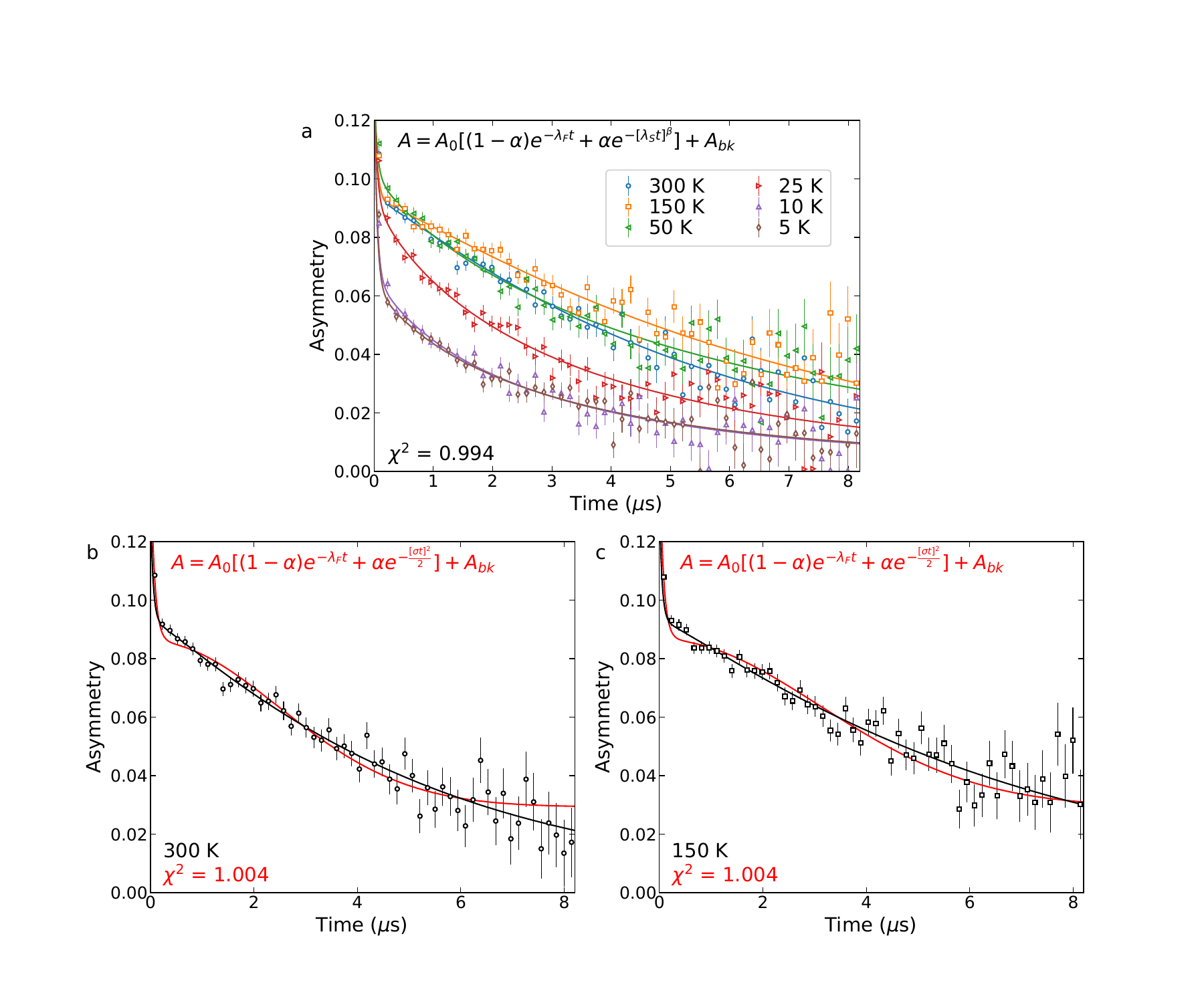}
    \caption{\justifying ZF asymmetry vs. time for the LCNO film across all measured temperatures, with theoretical fits to the equation shown. (b) Comparison between the best-fit (black line) and Gaussian fit (red line) for the 300 K dataset. (c) Comparison between the best-fit (black line) and Gaussian fit (red line) for the 150 K dataset. Measurements at 1 keV. Error bars represent $\pm$1 standard deviation. Reduced $\chi^2$ values are displayed for the fits shown.}
    \label{fig:ZF_LCNO_Full}
\end{figure*}

Figure \ref{fig:ZF_LCNO_Full}(a) shows the ZF asymmetry vs. time for the LCNO sample across the full set of measured temperatures. Here, we fit the data using a stretching parameter as discussed in equation 1 of the main text, to obtain the $\beta$ parameters shown in the main text. Interestingly, the asymmetry at long times ($\geq$ 6 $\mu$s) follows a non-monotonic trend with temperature. Specifically, the 150 K and 50 K asymmetry at the longest times is above that of the 300 K data, while the 25 K, 10 K, and 5 K data all show a reduction in asymmetry at long times, coincident with an abrupt decrease in the initial asymmetry. In Fowlie \textit{et al.}, it was noted that both (Nd,Sr)NiO$_2$ (NSNO) and (La,Sr)NiO$_2$ (LSNO) displayed an increase in ZF asymmetry at long times for the lowest temperature, which was taken as evidence of a static ordered magnetic state emerging \cite{FowlieMagnetism2022}.

As Gaussian ZF asymmetry shapes ($\beta \approx$ 2) have been reported for superconducting nickelates at the highest temperatures for NSNO, LSNO, (Pr,Sr)NiO$_2$ (PSNO), and LaNiO$_2$ (LNO) films on STO, we evaluate the appropriateness of fixing the $\beta$ parameter to equal 2. Figures \ref{fig:ZF_LCNO_Full}(b) and (c) show the best-fit (black) with variable $\beta$ and the Gaussian fit (red) for the 300 K and 150 K measurements. It can clearly be seen that forcing $\beta = 2$ yields a singificantly worse fit to both datasets, especially at low time. We therefore conclude that a Gaussian relaxation rate, associated with purely nuclear interactions, is not an appropriate description of the data.

\begin{figure*}
    \centering
    \includegraphics[width=0.9\textwidth]{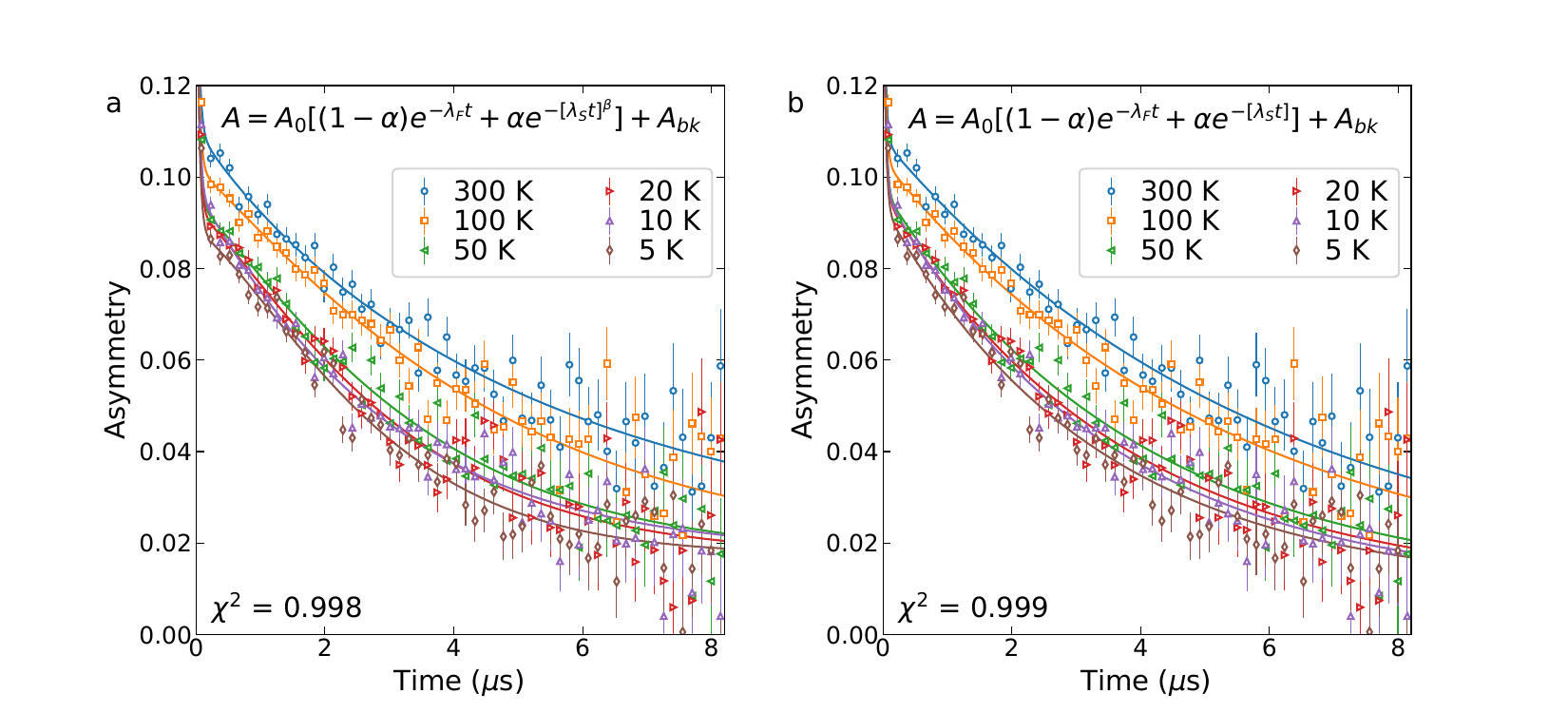}
    \caption{\justifying(a) ZF asymmetry vs. time for the NGO mosaic with theoretical fits allowing for variable stretching parameter ($\beta$). (b) ZF asymmetry vs. time for the NGO mosaic with theoretical fits constraining $\beta$ = 1. Measurements at 1 keV. Error bars represent $\pm$1 standard deviation. Reduced $\chi^2$ values are displayed for the fits shown.}
    \label{fig:ZF_NGO_Full}
\end{figure*}

We next examine, in Fig. \ref{fig:ZF_NGO_Full}, the full ZF asymmetry vs. time dataset for the NGO moasaic at a muon implantation energy of 1 keV. In Fig.  \ref{fig:ZF_NGO_Full}(a), we fit the data with a variable stretching parameter ($\beta$), while  Fig. \ref{fig:ZF_NGO_Full}(b) fits the data fixing $\beta = 1$. The two analyses yield largely indistinguishable curves with $\chi^2$s of 0.998 and 0.999, such that there is little statistical evidence of a departure from pure paramagnetic behavior in the NGO. Unlike the LCNO sample, the asymmetry at long time exhibits a monotonic temperature dependence, either decreasing or remaining constant with decreasing temperature.

\begin{figure*}
    \centering
    \includegraphics[width=0.9\textwidth]{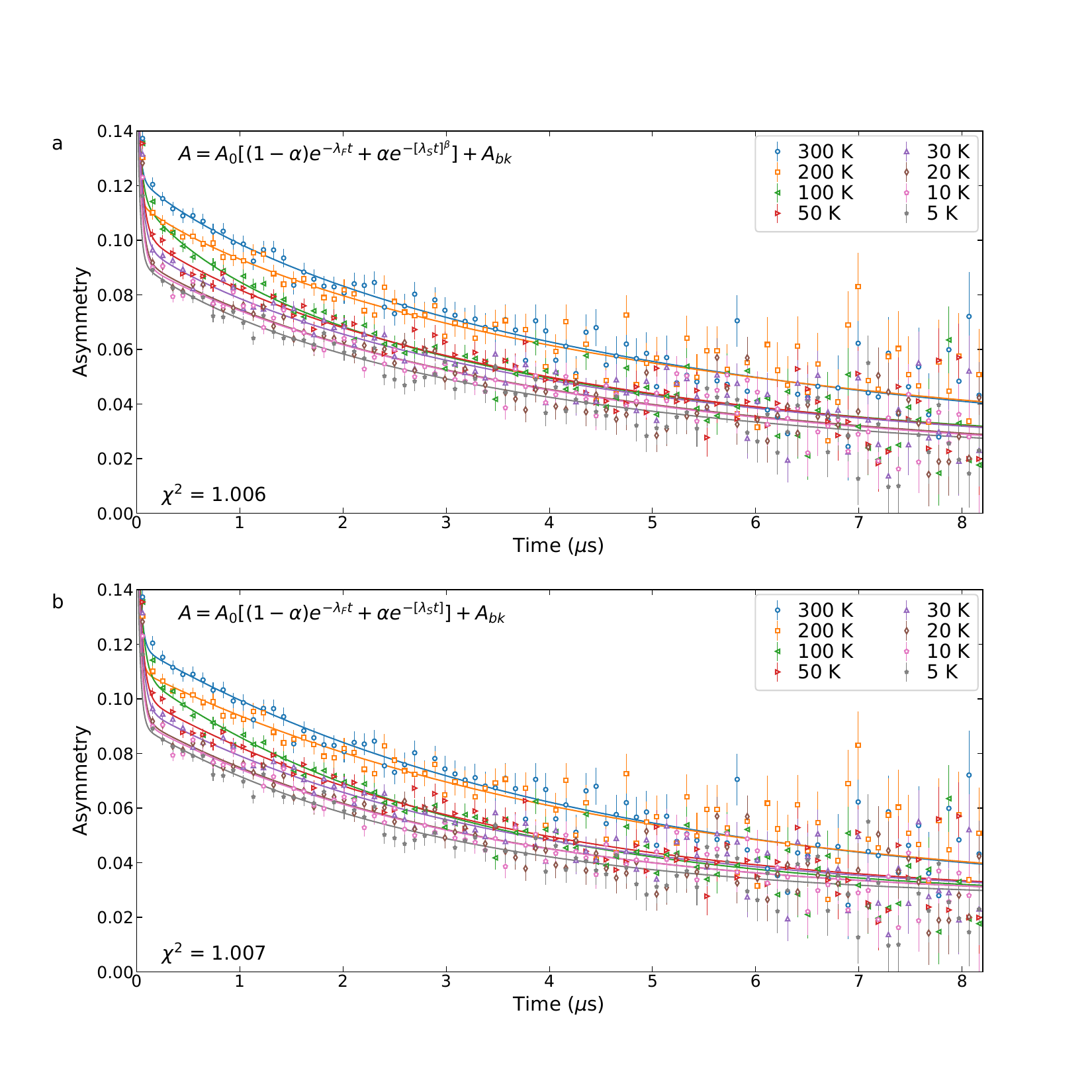}
    \caption{\justifying(a) ZF asymmetry vs. time for the SECNO mosaic with theoretical fits allowing for variable stretching parameter ($\beta$). (b) ZF asymmetry vs. time for the SECNO mosaic with theoretical fits constraining $\beta$ = 1. Measurements at 1 keV. Error bars represent $\pm$1 standard deviation. Reduced $\chi^2$ values are displayed for the fits shown.}
    \label{fig:ZF_SEC_Full}
\end{figure*}

\begin{figure*}
    \centering
    \includegraphics[width=0.9\textwidth]{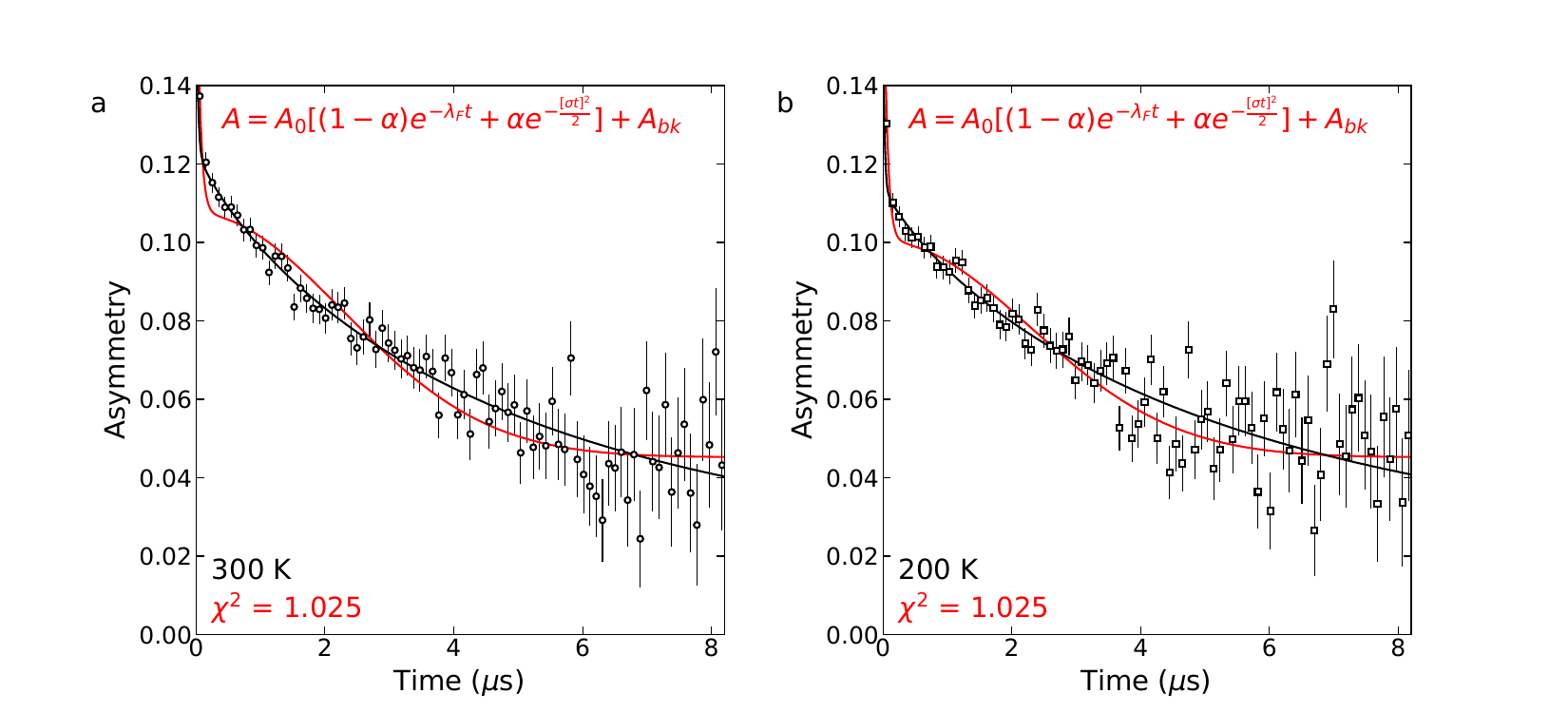}
    \caption{\justifying ZF asymmetry vs. time for the SECNO mosaic with theoretical fits allowing for variable $\beta$ (black line) and forcing $\beta = 2$ (red line) at (a) 300 K and (b) 200 K. Data taken using 1 keV muon implantation energy. Error bars represent $\pm$1 standard deviation. Reduced $\chi^2$ values are displayed for the Gaussian fits shown.}
    \label{fig:ZF_SEC_Gauss}
\end{figure*}

In Fig. \ref{fig:ZF_SEC_Full}, the full ZF asymmetry vs. time dataset is shown for the SECNO mosaic at a muon implantation energy of 1 keV. In Fig.  \ref{fig:ZF_SEC_Full}(a), we fit the data with a variable stretching parameter ($\beta$), while  Fig. \ref{fig:ZF_SEC_Full}(b) fits the data fixing $\beta = 1$. The resulting fits are very similar, supporting the conclusion in the main text that $\beta$ is near 1 and largely temperature-independent. We further examine the appropriateness of fitting the high-temperature ZF asymmetry for SECNO with a Gaussian lineshape in Fig. \ref{fig:ZF_SEC_Gauss}(a) and \ref{fig:ZF_SEC_Gauss}(b) for 300 K and 200 K, respectively. It is very clear that a Gaussian fit yields a significantly worse description of the data and should not be used.

\begin{figure*}
    \centering
    \includegraphics[width=0.9\textwidth]{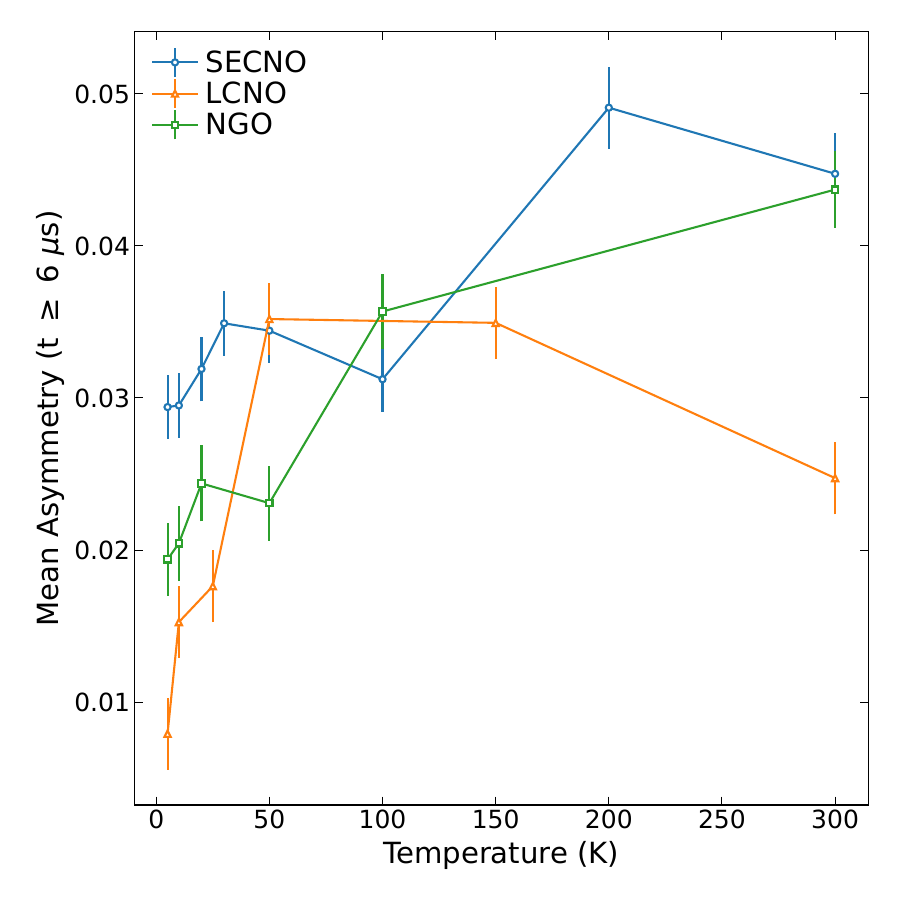}
    \caption{\justifying Temperature-dependent mean ZF asymmetry for times greater than 6 $\mu$s for 1 keV muon implantation energy across SECNO, LCNO, and NGO substrates. Error bars represent $\pm$1 standard deviation.}
    \label{fig:ZF_Longtime}
\end{figure*}

Having discussed all of the ZF data, we examine the asymmetry at long times in more detail. Figure \ref{fig:ZF_Longtime} shows the mean asymmetry for times equal to or exceeding 6 $\mu$s vs. temperature. As discussed above, the LCNO ZF asymmetry at t $\geq$ 6 $\mu$s exhibits non-monotonic behavior, increasing from 300 K to 150 K, then decreasing sharply from 50 K to 5 K. NGO monotonically decreases (within uncertainty) from 300 K to 5 K. The SECNO mosaic, on the other hand, is steady from 300 K to 200 K, decreases significantly between 200 K and 100 K, and then remains relatively steady from 100 K to 5 K. 

Lastly, we discuss tests of the statistical significance in certain features of the ZF SECNO data. In the main text, Figure 2(e), it may be noted that there is a slight trend towards increasing $\beta$ with decreasing temperature. Fitting $\beta$ for SECNO with a linear temperature function, $\beta$ = aT + b, and performing a Markov chain Monte-Carlo (MCMC) uncertainty analysis using the DREAM algorithm of the BUMPS python package, we find that a = -0.0002 K$^{-1}$ $\pm$ 0.0003  K$^{-1}$. That is, no statistically significant trend is observed. 

Similarly, the ZF $\lambda_{\rm S}$ vs. temperature for SECNO displays an apparent single-point peak at 100 K in Fig. 2(f) of the main text. Such a feature is plausible, as the ZF relaxation rate often spikes at magnetic transition temperatures, and 100 K is near the middle of the SECNO magnetic transition detected in wTF measurements. However, MCMC analysis indicates that this peak is not statistically significant. Fitting the data to a Gaussian shape on a linear background yields a peak amplitude which overlaps with zero within the 95\% confidence limit. The overall trend of increasing $\lambda_{\rm S}$ with decreasing temperature, however, is statistically significant above the 95\% confidence level.

\subsubsection{Note on the influence of substrates in LE$\mu$SR}

In LE$\mu$SR measurements of superconducting nickelate films, a principle challenge is the relatively thin nature of the samples. It is well known that the quality of superconducting nickelate films degrades dramatically with thickness, and it has been suggested that the superconductivity may be confined to a subsection of higher-quality material near the film/substrate interface. Consequently, we have confined our LE$\mu$SR experiments to samples which are less than 10 nm thick. We selected relatively thin capping layers to facilitate reduction. Consequently, our measurements were necessarily confined primarily to 1 keV muon implantation energies at which a significant background is expected from reflected muons.

On the other hand, Fowlie \textit{et al.} chose a geometry in which a much thicker capping layer was used in order to avoid the complexity of reflected muons \cite{FowlieMagnetism2022}. However, it is generally the case that the width of the muon distribution tracks the mean implantation depth, such that a thicker STO cap necessarily results in more muons in the STO substrate or STO cap. Thus, the fraction of muons implanted into the superconducting samples is less than 50\%, and, in the case of NSNO, less than 40\%.

The influence of the substrate on the resulting data is therefore an important question, especially given the observation that many significant features in the wTF data occur across the known cubic-tetragonal structure transition of STO, or below the muonium formation temperatures of 70 K. The STO exhibits sharp features in F$_M$ at these temperatures. The differences in high-temperature ZF asymmetry lineshape, in which Fowlie \textit{et al.} observe Gaussian relaxation while our observations are consistent with paramagnetism, are potentially attributable to either the difference between a diamagnetic and paramagnetic substrate or the effect of Eu dopants. More investigations will be required to resolve this question.

\subsection{Polarized Neutron Reflectometry Analysis}

Polarized neutron reflectometry (PNR) is a reciprocal space technique, in which the measured reflectometry curve encodes information about the real space sample geometry. This information must be extracted by means of fitting the data to a theoretical model. However, this model is never a unique solution to the neutron reflectometry data, such that all known information must be encoded into the model to constrain the number of potential solutions. In general, the simplest model which satisfies the known priors (encoded as model choices and fitting ranges) is preferred. An example of prior information might be the knowledge that NGO is a paramagnet, such that the magnetization across all three temperature/field combinations can be represented with a 1/T dependence, and that the magnetization must be linearly dependent on the field. We have implemented this constraint.

In order to demonstrate sensitivity to the physics of interest, it is critical to test alternative, counterfactual models for agreement with the data. We therefore present below two reasonable alternative models. These models are:\\
\textbf{Model A:} Best fit presented in the main text. Two SECNO sublayers with identical structures but different net in-plane magnetizations\\
\textbf{Model B:} Same as model A, but with a single uniform magnetization within the SECNO layer\\
\textbf{Model C:} Same as model A, but all magnetism is confined to the paramagnetic $\mathrm{NGO}$ substrate\\

All models have been refined in the same manner, with the Refl1D software program using the DREAM algorithm implemented in the bumps Python package \cite{Refl1D, kirby2012phase}. Uncertainties are derived from MCMC sampling based on $10^6$ samples taken after model convergence. As shown below, Model B yields a significantly worse description of the spin splitting when compared to Model A in the main text. In particular, the positive low-$Q_Z$ feature in the spin asymmetry is shifted to lower $Q$ and does not well-describe the long positive tail in the 35\,K, 1\,T dataset which stretches from $Q = 0.25 \,\mathrm{nm^{-1}}$ to $Q_Z = 0.70 \,\mathrm{nm^{-1}}$. The fit to the 9\,K, 1\,T data is similarly problematic, and also undershoots the magnitude of the feature near $Q_Z = 0.35\,\mathrm{nm^{-1}}$. Consequently, this model yields a higher $\chi^2$ and is rejected, although it is still useful in identifying the SECNO layer as magnetic and yields similar magnetizations for the two high-field datasets while identifying the magnetization at 9\,K, 10\,mT as essentially zero. Model C is substantially worse than Model B, being completely unable to describe the spin-splitting in any portion of the $Q$-range. We therefore conclude that a net magnetization must reside in the SECNO layer.

\begin{figure*}
    \centering
    \includegraphics[width=0.9\textwidth]{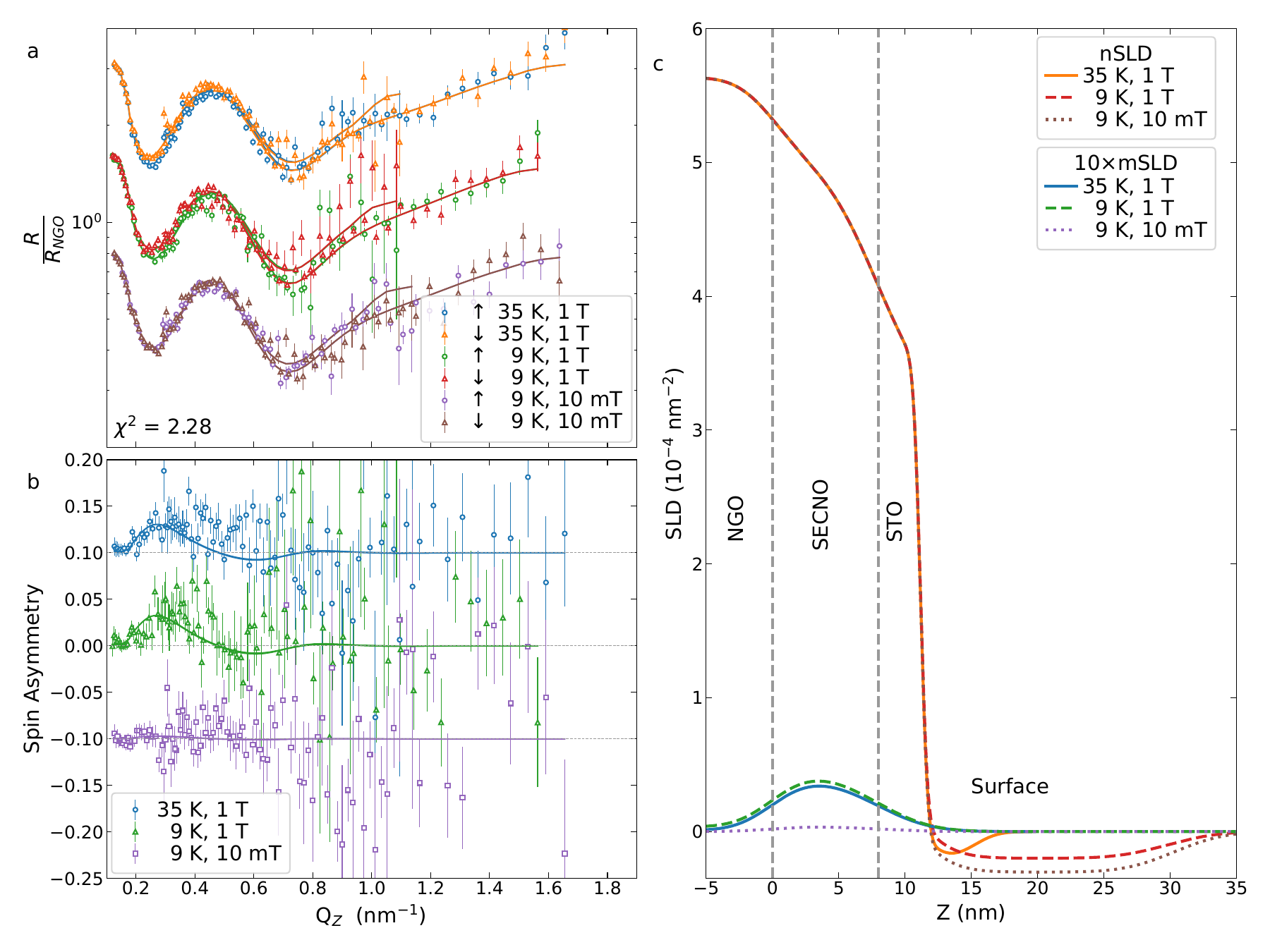}
    \caption{\justifying(a) Spin-dependent neutron reflectivity, normalized by theoretical substrate reflectivity vs.\ $Q_Z$ alongside theoretical fits from Model B. Curves offset for visual clarity. (b) Spin asymmetry vs.\ $Q_Z$ calculated from the data in (a), alongside theoretical curves from Model B. Curves offset for visual clarity. (c) Best-fit nuclear and magnetic neutron scattering length densities from Model B vs.\ distance from the film/substrate interface ($Z$), based on Model A with two distinct magnetic regions in the SECNO. Error bars represent $\pm 1$ standard deviation.}
    \label{fig:PNRModelB}
\end{figure*}

\begin{figure*}
    \centering
    \includegraphics[width=0.9\textwidth]{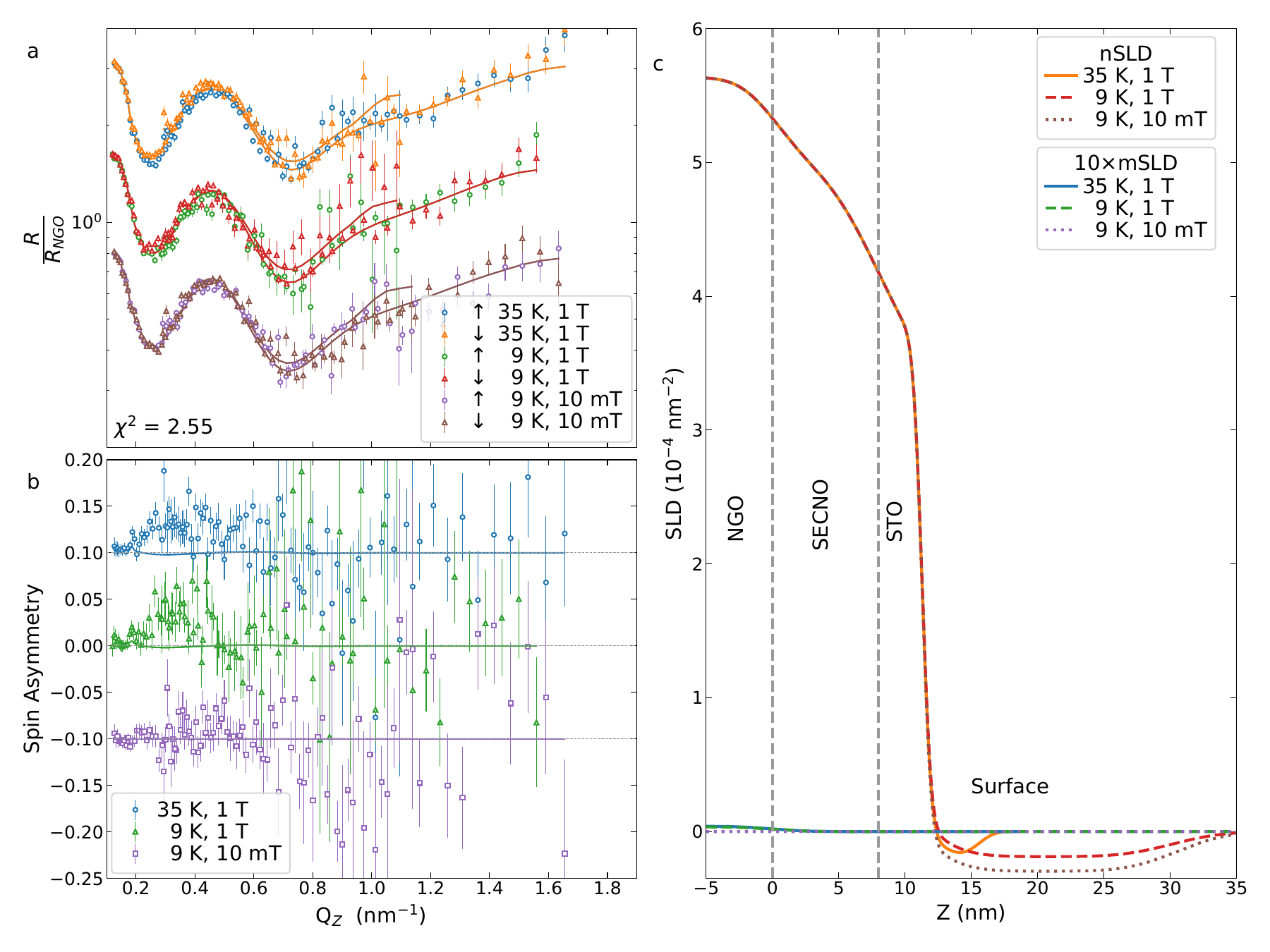}
    \caption{\justifying(a) Spin-dependent neutron reflectivity, normalized by theoretical substrate reflectivity, vs.\ $Q_Z$, alongside theoretical fits from Model C. Curves offset for visual clarity. (b) Spin asymmetry vs.\ $Q_Z$ calculated from the data in (a), alongside theoretical curves from Model C. Curves offset for visual clarity. (c) Best-fit nuclear and magnetic neutron scattering length densities from Model C vs.\ distance from the film/substrate interface ($Z$), based on Model A with two distinct magnetic regions in the SECNO. Error bars represent $\pm 1$ standard deviation.}
    \label{fig:PNRModelC}
\end{figure*}

\subsection{LCNO Transport}

\begin{figure*}
    \centering
    \includegraphics[width=0.9\textwidth]{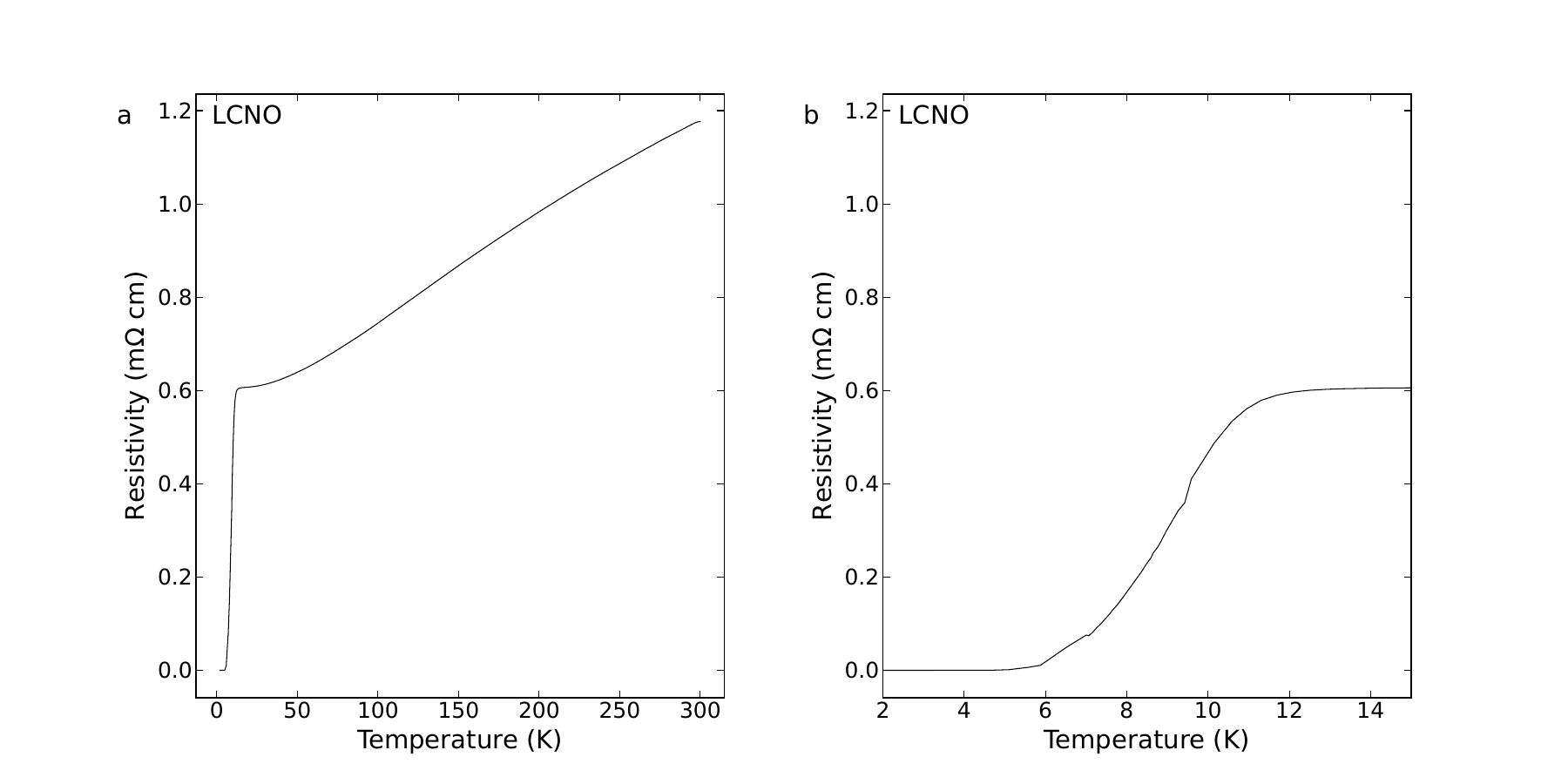}
    \caption{\justifying(a) Resistivity vs.\ temperature for the LCNO sample, full temperature range. (b) Resistivity vs.\ temperature for the LCNO sample, zoomed in view of the superconducting transition.}
    \label{fig:LCNO_Transport}
\end{figure*}

Figure \ref{fig:LCNO_Transport} shows resistivity vs.\ temperature for the LCNO measured by LE$\mu$SR. The entire range is shown in \ref{fig:LCNO_Transport}(a), with an expanded view of the superconducting transition in \ref{fig:LCNO_Transport}(b).

\section*{References}
\bibliography{references}



\end{document}